\documentclass[a4paper,12pt]{elsarticle}
\usepackage[a4paper]{geometry}
\usepackage[cm]{fullpage}
\usepackage{amsthm,amsmath,amssymb}
\usepackage{graphicx}
\usepackage{url}
\usepackage{natbib}
\usepackage{hyperref}

\numberwithin{equation}{section}
\theoremstyle{plain}
\newtheorem{theorem}{Theorem}[section]

\theoremstyle{definition}

\newtheorem{example}[theorem]{Example}
\theoremstyle{remark}
\newtheorem{remark}[theorem]{Remark}
\numberwithin{equation}{section}

\newcommand{\spc}{\mathbb R}
\newcommand{\spctim}{\spc \times [0,\infty)}

\newcommand{\1}{\mathbf 1}

\newcommand{\del}{\partial}

\begin{document}

\begin{frontmatter}

\title{Variable Order Fractional Fokker-Planck Equations derived from 
Continuous Time Random Walks}
\author[UNSW]{Peter Straka}
\ead{p.straka@unsw.edu.au}
\address[UNSW]{School of Mathematics \& Statistics, UNSW Sydney, Sydney NSW 2052, Australia}

\begin{abstract}
Continuous Time Random Walk models (CTRW) of anomalous diffusion are 
studied, 
where the anomalous exponent $\beta(x) \in (0,1)$ varies in space.  This 
type of situation occurs e.g.\ in biophysics, where the density of the 
intracellular matrix varies throughout a cell.  Scaling limits of CTRWs
are known to have probability distributions which solve fractional 
Fokker-Planck type equations (FFPE).  This correspondence between 
stochastic processes and FFPE solutions has many useful extensions e.g.\ 
to nonlinear particle interactions and reactions, but has not yet been
sufficiently developed for FFPEs of the``variable order'' type with non-constant 
$\beta(x)$. 

In this article, variable order FFPEs (VOFFPE) are derived from scaling 
limits of CTRWs.  The key mathematical tool is the 1-1 correspondence 
of a CTRW scaling limit to a bivariate Langevin process, which tracks the
cumulative sum of jumps in one component and the cumulative sum of waiting 
times in the other. The spatially varying anomalous exponent is
modelled by spatially varying $\beta(x)$-stable L\'evy noise in the 
waiting time component. 
The VOFFPE displays a spatially heterogeneous temporal scaling behaviour, 
with generalized diffusivity and drift coefficients whose units are 
length$^2$/time$^{\beta(x)}$ resp.\ length/time$^{\beta(x)}$. 
A global change of the time scale results in a spatially varying change in 
diffusivity and drift. 

A consequence of the mathematical derivation of a VOFFPE from CTRW limits 
in this article is that a solution of a VOFFPE can be approximated via 
Monte Carlo simulations.  
Based on such simulations, we are able to confirm that the VOFFPE is consistent under a 
change of the global time scale. 
\end{abstract}


\end{frontmatter}

\section{Introduction}

Subdiffusive processes are characterized by a sublinearly growing mean squared
displacement proportional to $t^\beta$, $0 < \beta < 1$, and have been reported
in many experimental systems
\cite{Metzler2000,TMT04,Santamaria2006a,Banks2005,Regner2013}.
For a majority of these systems, long rests of a walker are thought to be the
main mechanism causing subdiffusion, and Continuous Time Random Walks (CTRWs)
with heavy-tailed waiting times effectively capture this phenomenon
\cite{Scher1975,BG1990,Metzler2000}.  CTRWs have become a widely used model,
particularly because scaling limits of their probability distributions can be
modelled by fractional differential equations and Fokker-Planck equations
\cite{BMK00,Baeumer2001,HLS10PRL,Hahn11} which can be extended tractably to
model particle reactions \cite{Langlands2008d,Angstmann2013} and nonlinear
interaction effects \cite{StrakaFedotov14}.

In the majority of the literature on CTRWs and FFPEs (fractional Fokker-Planck
equations), the
fractional parameter $0 < \beta < 1$ is treated as a global constant.  The
situation where $\beta$ varies in space, however, is of interest in physics,
since the strength of a trapping effect may vary throughout a disordered
medium
\cite{Wong04,Korabel2010,Stickler2011,Fedotov2012,StrakaFedotov14}.
The experiments in \cite{Wong04} find that for the diffusion of a bead in
F-actin networks, the fractional parameter depends solely on the ratio of
bead size to mesh width, and hence a physical system with continuously varying
fractional parameter can be designed.

There are various works in the statistical physics and computational literature
that consider ``variable
order'' fractional subdiffusion equations (see e.g.\ 
\cite{Chechkin2005a,Sun2009,Chen2010,Sandev2015} and
the references therein). 
We highlight three such variable variable order FFPE (VOFFPE) below:
\begin{description}
  \item [\cite{Chechkin2005a}] Equation (11), the original derivation of a 
  VOFFPE from CTRWs: 
  \begin{align} \label{eq:Chechkin}
    \frac{\partial}{\partial t} P(x,t) 
    = \frac{\partial}{\partial x^2} 
    \left[K(x) D_t^{1-\beta(x)} P(x,t)\right]
  \end{align}
  \item [\cite{Sun2009}] Equation (8): 
  \begin{align}
  \label{eq:wrong1}
    D_t^{\beta(x)} P(x,t) &= K \frac{\partial^2 P(x,t)}{\partial x^2}
  \end{align}
  \item [\cite{Chen2010}] Equation (1.2):
  \begin{align}
  \label{eq:wrong2}
  \frac{\partial P(x,t)}{\partial t} &= K {_0 D_t^{1-\beta(x)}}
  \frac{\partial^2 P(x,t)}{\partial x^2} + f(x,t)
  \end{align}
\end{description}
In \eqref{eq:wrong1}, the fractional temporal derivative is of Caputo type,
whereas in \eqref{eq:wrong2}, it is of Riemann-Liouville type.
We note the following issues in the applicability of these three prototypes:
\begin{itemize}
  \item
  In the above works, the consistency of physical dimensions is not addressed: 
  An application of the
  fractional derivative of order $\beta$ is dimensionally equivalent to a
  division by time$^{\beta}$, and it appears that one side of the equation has 
  integer powers of time whereas the other side has a fractional power of time.
  A non-dimensional interpretation of the differential equation seems to be
  necessary, but it is unclear how time scale parameters enter into the
  equation, or how dynamics are related at two different time scales.
  \item
  Since the operators $\partial/\partial x$ and $\partial^2 / \partial x^2$
  do not commute with the variable order fractional derivative 
  $D_t^{1-\beta(x)}$, the three equations have intrinsically different solutions
  $P(x,t)$. 
  Only Equation \eqref{eq:Chechkin}
  has been derived from a CTRW scaling limit, which means that 
  Equations \eqref{eq:wrong1} and \eqref{eq:wrong2} are
  disadvantageous, because i) it is not immediately clear if solutions are 
  positive,
  ii) probability densities cannot be computed via Monte Carlo simulations,
  and iii) extensions for chemical reactions or other particle interaction are 
  ad hoc in the sense that they are not based on a particle model.
  \item
  The three works above do not address the issue of how to model space- and 
  time-dependent drift and diffusivity, though this is useful for the modelling 
  of e.g.\ chemotaxis and chemokinesis \cite{StrakaFedotov14}.
\end{itemize}
In this article, we resolve the above issues: 
From CTRW limit dynamics, we derive the VOFFPE \eqref{eq:FFPE-with-scale}, 
as a logical extension of \eqref{eq:Chechkin}. 
We show that our VOFFPE is dimensionally consistent, and elucidate consequences 
of the spatially inhomogeneous temporal scaling. 
Our framework is very general, and rests on only one unrestrictive assumption: 
The \emph{Semi-Markov property}, which states that the next waiting time of a 
CTRW is independent of the current time, and depends only on the current 
position in space \cite{Germano2009,Meerschaert2014}\footnote{But the next jump 
may depend on both location and current time.}.
This framework allows for a space- and time-dependent diffusivity and drift,
for which we detail conditions guaranteeing existence and uniqueness of a 
solution to the VOFFPE \eqref{eq:FFPE-with-scale}.

There is a 1-1 correspondence between Semi-Markov processes and 
\emph{Markov additive processes} (MAP) \cite{cinlar1974markov}.
These are Markov processes in space-time, with the property that
the future trajectory is independent of the past, given the present 
\emph{space} component \emph{only} (as opposed to given the present value 
of \emph{both} components). 
We fully exploit this 1-1 correspondence and 
represent CTRWs and their scaling limits as Langevin processes in 
space-time $\spctim$.
In the Langevin setting, spatially varying fractional parameters 
for the waiting time distribution can be introduced relatively easily, 
which is a useful tool in the derivation of the VOFFPE.

A recent result in \cite{BaeumerStraka16} then provides the
generalized FFPE (Fractional Fokker-Planck Equation) \eqref{eq:FFPE}
which governs a CTRW scaling limit $X_t$ in the sense that its unique 
solution $P(x,t)$ is the probability distribution of $X_t$. 
The generalized FFPE \eqref{eq:FFPE} contains a 
space-dependent memory kernel $V(x,s)$ which admits an interesting 
interpretation
based on renewal theory.  Moreover, if the dynamics are locally fractional
with parameter $\beta(x)$, the VOFFPE
\eqref{eq:FFPE-with-scale} results from the generalized FFPE
\eqref{eq:FFPE}.

Section 2 describes the underlying dynamics in space-time, where the first
component tracks the location of a particle and the second component the
physical time that has passed.  Scaling limits are identified as bivariate
Langevin processes which uniquely correspond to CTRW limits.
In Section 3 we utilize a result from the mathematics literature that links CTRW 
limits with fractional Fokker-Planck equations.
In Section 4 we show that the variable-order FFPE \eqref{eq:FFPE-with-scale} is consistent, by  calculating matching solutions at different time scales via a Monte-Carlo simulation of CTRW trajectories.

\section{The space-time setting}

\subsection{CTRWs are space-time Markov Chains}

The trajectory of a CTRW can be reduced to a series of points in
space-time, without loss of information: If $t_1, t_2, \ldots$
denote the jump times of a CTRW and $x_1, x_2, \ldots$ the locations
of the end points of the jumps, then the location of the CTRW at time
$t$ is
\begin{align} \label{eq:CTRW-trajectory}
X^{(c)}_t = x_k \text{ for } k \ge 0 \text{ such that }
t_k \le t < t_{k+1}.
\end{align}
We write $c$ for a CTRW scaling parameter, as will be detailed below.
Moreover, write $t_0$ for the starting time and $x_0$ for the starting
location. Then the pair sequence
$(t_0, x_0), (t_1, x_1), (t_2, x_2), \ldots$
must define a Markov chain in space-time $\spctim$,
by the Semi-Markov assumption that the $k+1$-th waiting time $t_{k+1} - t_k$
and the $k+1$-th jump $x_{k+1} - x_k$ are both independent of the past
$(t_0, x_0), \ldots, (t_{k-1}, x_{k-1})$ given the current $(t_k, x_k)$.

The distribution of the Markov chain $(t_k, x_k)$ is thus uniquely
determined by the initial location $(x_0, t_0)$ in space-time,
and a probability kernel on space-time
\begin{align} \label{eq:STJK}
K^{(c)}(y,w | x,s)
\end{align}
denoting the probability distribution of the next jump 
$y = x_{k+1} - x_k$ after the next waiting time
$w = t_{k+1} - t_k$, conditional on $(x_k, t_k) = (x,s)$.
As an example, let $(t_0, x_0) = (0,0)$ and
\begin{align} \label{eq:STJK-example}
K^{(c)}(y,w | x,s) = \left[\frac{1}{2} \delta_{+\chi}(y) + \frac{1}{2} \delta_{-\chi}(y)\right] \beta(x) (1+w/\tau)^{-1-\beta(x)} \, dw,
\quad w > 0, \quad \beta(x) \in (0,1).
\end{align}
Then \eqref{eq:STJK-example} defines a CTRW with Pareto waiting times at scale
$\tau$ and jumps $\pm \chi$ with equal probability $1/2$.  The tail parameter 
$\beta$ varies as a function $\beta(x)$ in space.

We will view $c > 0$ as a ``master'' scaling parameter, 
in the sense that as $c \uparrow \infty$, both $\chi \downarrow 0$ and 
$\tau \downarrow 0$. 
Thus for every $c$ we have defined a Markov chain on $\spctim$, which
in turn defines a CTRW at scale $\chi$ in space and $\tau$ in time
via \eqref{eq:CTRW-trajectory}.
We are interested in the limiting CTRW process as $c \to \infty$ (if there 
is one).
Theorem \ref{theorem1} specifies the exact conditions on the rate at 
which $\chi \downarrow 0$ and $\tau \downarrow 0$, so that a limiting 
CTRW process exists. 
See Subsection \ref{sec:scaling} for more details.

\subsection{The limiting space-time Langevin Process}

We seek to identify and characterize a class of possible limit processes
$(Y_u, Z_u)$ of the above Markov chains,
in the scaling limit as $c \to \infty$. As it turns out, this class of 
limiting stochastic processes are solutions to
Langevin equations in space-time (or solutions to stochastic differential
equations with L\'evy noise, or ``jump-diffusions'').
We characterize this class of Langevin processes in this section, and detail 
a construction of these processes via Poisson Random Measures.
Then in Section \ref{subsec:convergence}, we utilize results from the theory of
convergence of stochastic processes to give exact conditions of convergence
to these Langevin processes.

\paragraph{Poisson Random Measure}
The standard construction of L\'evy noise is via Poisson Random Measures
\cite{Applebaum}.  We describe an approach which is amenable to a 
spatially varying coefficient $\beta(x)$. 
We first define the mean measure $M$ to be the Lebesgue measure on 
$(0,\infty) \times [0,\infty)$; the first component $(0,\infty)$ corresponds 
to the size of the jumps $w > 0$ of the temporal component $Z_u$, whereas $[0,\infty)$
corresponds to an auxiliary (or ``operational'') time, during which these jumps accumulate. 
Lebesgue measure is nothing but the surface measure, defined uniquely by the requirement 
that a rectangle $B = (a,b) \times (c,d) \subset (0,\infty) \times [0,\infty)$
have the measure $M(B) = (b-a)(d-c)$.  
A Poisson Random Measure with intensity measure $M$ then is a random measure 
$N(dw, ds) = \sum_{r \in \mathcal R} \delta_{r}(dw, ds)$ where $\delta_r$ 
denotes a Dirac point measure at the point $r \in (0,\infty) \times [0,\infty)$
and $\mathcal R \subset (0,\infty) \times [0,\infty)$ is a random discrete set of points 
such that the following two criteria hold: 
\begin{itemize}
  \item 
  For any bounded rectangle $B$, the number of 
  points of $\mathcal R$ that lie in $B$ is Poisson distributed with mean 
  $M(B)$.
  \item
  The point counts in disjoint rectangles are independent. 
\end{itemize}

\paragraph{L\'evy measure}
The points in $\mathcal R$ are transformed in their first component according 
to the \emph{coefficient function} $F$.  For instance, if we let 
\begin{align} \label{eq:coefficient-F}
  F(w) = \left( \Gamma(1-\beta) w \right)^{-1/\beta}, \quad 0 < \beta < 1
\end{align}
then it is not difficult to confirm that the set of points 
$F(\mathcal R) := \{(F(w), s): (w,s) \in \mathcal R\}$ 
define a Poisson Random Measure on $(0,\infty) \times [0,\infty)$ with mean 
measure 
\begin{align}
  M_F((a,b) \times (c,d)) = \int_a^b \nu(w)\,dw \times (d-c),
\end{align}
where 
\begin{align} \label{eq:stable-levy-measure}
\nu(w) = \frac{\beta}{\Gamma(1-\beta)} \, w^{-1-\beta}, \quad w > 0, \quad 
0 < \beta < 1.
\end{align}
The density $\nu$ satisfies\footnote{$a \wedge b := \min\{a, b\}$}
\begin{align}
\label{eq:Levy-condition}
\int_{w > 0}(1 \wedge w) \nu(w)\,dw < \infty,  
\end{align}
and is hence called a \emph{L\'evy measure};
the points in $F(\mathcal R)$ are referred to as \emph{L\'evy noise}.
Finally, the L\'evy process $Z_u$ is constructed from L\'evy noise via 
\begin{align} \label{eq:summed-noise}
  Z_u = \int\limits_{0 \le s \le u} \int\limits_{w > 0} F(w) N(dw,ds)
  := \sum_{(w,s) \in \mathcal R:\, s \le u} F(w).
\end{align}
Note that condition \eqref{eq:Levy-condition} guarantees the convergence of the 
sum in \eqref{eq:summed-noise}. 
Increasing L\'evy processes are called subordinators in the mathematics
literature.  If $F$ is chosen as in \eqref{eq:coefficient-F}, then the above 
subordinator has stable marginals, i.e.\ its Laplace 
transforms are \cite{Bertoin04}
\begin{align}
  \langle \exp(-\lambda Z_u) \rangle = \exp(-u \lambda^\beta), \quad u \ge 0.
\end{align}

\paragraph{Tempered $\beta$-stable subordinator}
A tempered $\beta$-stable subordinator has a L\'evy measure of the form
\begin{align}
  \label{eq:tempered-stable-levy-measure}
  \nu(w) = \frac{\beta}{\Gamma(1-\beta)}\, w^{-1-\beta}\,e^{-\theta w}\,dw, 
  \quad w > 0, \quad 0 < \beta < 1, \quad \theta \ge 0. 
\end{align}
Here $\theta$ is called the tempering parameter. A positive $\theta$ decreases 
the occurrence of very large jumps $w$ of $Z_u$, and the mean of $Z_u$ becomes 
finite.  If instead of \eqref{eq:coefficient-F} we let 
\begin{align} \label{eq:general-F}
  F(w) = \inf\left\lbrace z > 0: \overline \nu(z) \le w \right\rbrace
\end{align}
where $\overline \nu(z) = \int_z^\infty \nu(w')\,dw'$ and $\nu$ is as in \eqref{eq:tempered-stable-levy-measure}, then 
\eqref{eq:summed-noise}
yields a tempered stable subordinator $Z_u$.

\paragraph{Varying noise}
We now assume that the coefficients $\beta$ and $\theta$ of the tempered 
$\beta$-stable subordinator are functions of $x$, the location in space. 
To this end, we make 
\eqref{eq:tempered-stable-levy-measure} space-dependent via
\begin{align}
  \label{eq:varying-tempered-stable-levy-measure}
  \nu(w | x) = \frac{\beta(x)}{\Gamma(1-\beta(x))}\, 
  w^{-1-\beta(x)}\,e^{-\theta(x) w}\,dw, 
  \quad w > 0, \quad 0 < \beta(x) < 1, \quad \theta(x) \ge 0, 
\end{align}
where we assume that $\beta(x) \in (0,1)$ and $\theta(x) \ge 0$ are  
functions of $x$.
We also make the coefficient function \eqref{eq:general-F} space-dependent via
\begin{align} \label{eq:varying-general-F}
  F(w|x) = \inf\left\lbrace z > 0: \overline \nu(z | x) \le w 
  \right\rbrace,
\end{align}
where $\overline \nu(z | x) = \int_z^\infty \nu(w'|x)\,dw'$.
Then $Z_u$ as defined in \eqref{eq:summed-noise} becomes a tempered stable 
subordinator with variable parameters $\beta(x)$ and $\theta(x)$. 
We have now set up the necessary tools for the limiting Langevin 
dynamics $(Y_u, Z_u)$ in space-time.

\paragraph{Space-time Langevin Process}
Before we discuss the stochastic process scaling limit mechanism, we 
define the class of scaling limits $(Y_u, Z_u)$ which is suitable for our CTRW setting. This class consists of all Langevin processes in 
$\spctim$ of the form
\begin{align} \label{eq:SDEY}
dY_u &= b(Y_{u}, Z_{u})\,du + \sqrt{a(Y_{u}, Z_{u})}\, dW_u \\
\label{eq:SDEZ}
dZ_u &= d(Y_{u})\,du + \int_{w > 0} F(w | Y_{u}) \, N(dw, du).
\end{align}
Here $Y_u$ denotes location in space; $Z_u$ the corresponding (physical) time;
$u$ denotes auxiliary time, resulting from the rescaled number of CTRW steps;
$W_u$ is an independent Brownian motion; 
$b$ and $a$ are space- and time-dependent
spatial drift resp.\ diffusivity functions; 
$d \ge 0$ is a temporal drift coefficient function;
and $F$ is as in \eqref{eq:varying-general-F}.

The distinction between physical time $Z_u$ and auxiliary time $u$ is helpful
as it allows to maintain a coherent physical interpretation of the FFPE derived
in Section \ref{sec:FFPE}.
The coefficient functions $b(y,t)$ and $a(y,t)$ measure
length (resp.\ length squared) \emph{per unit auxiliary time}.  
Similarly,
$d$ measures physical time per unit auxiliary time.
$F$ is 
dimensionless, while $N(dw,du)$ counts the point process points
per auxiliary time, and hence its two-dimensional integral over both physical 
time $w$ and auxiliary time $u$ is physical time valued.
Equation \eqref{eq:SDEY} has already appeared in \cite{Weron2008}, where
it was assumed that $Z_u$ is a homogeneous $\beta$-stable subordinator (L\'evy
flight).
The inhomogeneous form of $Z_u$ in \eqref{eq:SDEZ} is key to the generalization
to CTRW scaling limits with spatially inhomogeneous waiting times.

\paragraph{Conditions on coefficient functions}
If the coefficient functions $a, b, d$ and $F$ satisfy certain local 
Lipschitz and growth conditions \cite[Chapter 6]{Applebaum}, then the bivariate
Langevin equation \eqref{eq:SDEY} -- \eqref{eq:SDEZ} has a unique stochastic 
process solution.  Specifying $F(w|x)$ to the form \eqref{eq:varying-general-F} which depends on 
$\beta(x)$ and $\theta(x)$, we can employ a result by \cite{Tsuchiya1992}, 
which states that the Lipschitz and growth conditions are 
satisfied if we assume that the functions $a$, $b$, $d$, $\beta$ and $\theta$ 
are bounded and continuously differentiable with bounded derivative.
See Section~\ref{subsec:Lip-gro} for details.

\subsection{Limit theorem for the Langevin equation}
\label{subsec:convergence}
\paragraph{Infinitesimal generator}
The limiting Langevin stochastic processes $(Y_u,Z_u)$ are Markov
processes with the Feller property, which means that at any time $u$, their distribution depends 
continuously on their starting point $(x,s)$ (see the textbooks 
\cite{Kallenberg, Applebaum} for more details).
The distributions of their sample trajectories
are uniquely determined by their infinitesimal generator $\mathcal A$, 
which is the operator
\begin{align}
\mathcal A f(x,s) = \lim_{u \downarrow 0}
\left \langle f(Y_u, Z_u)\right \rangle_{(x,s)} / u, 
\quad f \in \text{dom}(\mathcal A).
\end{align}
Its domain $\text{dom}(\mathcal A)$ is the space of all real-valued, 
continuous functions vanishing at $\infty$ for which the above limit exists;
within this space, the twice continuously differentiable functions $f$ are 
dense, and hence it suffices to consider such $f$.
The angle brackets, $\langle \, \rangle_{(x,s)}$ denote the ensemble average over all
trajectories starting at $(x,s)$. 
The general form of $\mathcal A$ is given e.g.\ in 
\cite[Section 6.7.1]{Applebaum}. 
However, since $Z_u$ has increasing sample paths and $Y_u$ has 
continuous sample paths, $\mathcal A$ has the simpler form
\begin{align} \label{eq:inf-gen}
\begin{split}
\mathcal A f(x,s)
&= b(x,s) \frac{\del }{\del x} f(x,s)
+\frac{1}{2} a(x,s) \frac{\del^2}{\del x^2} f(x,s)\\
&+ d(x) \frac{\partial}{\partial s} f(x,s)
+ \int_0^\infty \left[f(x,s+w)-f(x,s)
\right] \nu(w|x)\, dw.
\end{split}
\end{align}
Note that by the Semi-Markov assumption, 
the distribution of waiting times depends on space only and not
on the current time; 
hence $d(x)$ and $\nu(w|x)$ only depend on $x$ and not on $s$.

\paragraph{Conditions for stochastic process convergence}
First, we embed the Markov chain with kernel \eqref{eq:STJK} from 
Section 2.1 into continuous time. Define a continuous time Markov 
chain $(Y^{(c)}_u, Z^{(c)}_u)$ on 
$\spctim$ as follows: at starting location $(x,s)$, wait an amount of time which is exponentially distributed with mean $1/c$. (This time is in ``auxiliary time'' $u$, and is not a CTRW waiting time!) Then jump to the location $(x+y, s+w)$, where the distribution of $(y,w)$ is determined by \eqref{eq:STJK}, and repeat. The resulting process has the infinitesimal generator
(see e.g.\ \cite[Proposition 17.2]{Kallenberg}) 
\begin{align} \label{eq:jump-process-generator}
\mathcal A^{(c)} f(x,s) = c \iint [f(x+y, s+w) - f(x,s)] K^{(c)}(y,w | x,s)\,dy\,dw.
\end{align}
A key result for Feller processes is that stochastic process convergence is equivalent to the convergence of infinitesimal generators \cite[Theorem 17.25]{Kallenberg}:
\begin{align} \label{eq:generator-condition}
\lim_{c \to \infty}(Y^{(c)}_u, Z^{(c)}_u) = (Y_u, Z_u)
\text{ if and only if }
\lim_{c \to \infty} \mathcal A^{(c)}f(x,s) = \mathcal A f(x,s), 
\quad f \in {\rm Dom}(\mathcal A).
\end{align}
The convergence on the left side of \eqref{eq:generator-condition} is in the sense of stochastic processes with respect to the Skorokhod $J_1$ topology, see e.g.\ \cite{Whitt2010}. This mode of convergence implies \emph{finite 
dimensional convergence}, i.e.\ the fact that for any collection
of times $u_1 < \ldots < u_n$, the convergence in distribution of the 
random vectors
\begin{align}
\left((Y^{(c)}_{u_1}, Z^{(c)}_{u_1}), \ldots, (Y^{(c)}_{u_n}, Z^{(c)}_{u_n}) \right) 
\to 
\left((Y_{u_1}, Z_{u_1}), \ldots, (Y_{u_n}, Z_{u_n}) \right)
\end{align}
holds. 

The two theorems below connect \eqref{eq:generator-condition} to the convergence of CTRWs: 

\begin{theorem} \label{theorem1}
The right side of \eqref{eq:generator-condition} holds if the following 
conditions all hold: 
\begin{align} \label{eq:cond1}
\lim_{\epsilon \downarrow 0} \lim_{c \to \infty}
c \iint\limits_{|z|< \epsilon,\,0<  w < \epsilon} z K^{(c)}(z,w | x,s)\,dz\,dw &= b(x,s)
\\ \label{eq:cond2}
\lim_{\epsilon \downarrow 0} \lim_{c \to \infty}
c \iint\limits_{|z|< \epsilon, \,0<w < \epsilon} z^2 K^{(c)}(z,w | x,s)\,dz\,dw &= a(x,s)
\\ \label{eq:cond3}
\lim_{\epsilon \downarrow 0} \lim_{c \to \infty}
c \iint\limits_{|z|< \epsilon, \,0<w < \epsilon} w K^{(c)}(z,w | x,s)\,dz\,dw &= d(x)
\\
\label{eq:cond4}
\lim_{c \to \infty}
c \iint\limits_{|z| \ge \varepsilon \text{ or } w \ge 0} g(z,w) K^{(c)} (z,w | x,s)\,dz\,dw &= \int g(0,w) \nu(w|x)\,dw, \quad \varepsilon > 0,
\end{align}
where $g(z,w)$ is any bounded continuous function which vanishes in a neighbourhood of the origin. 
\end{theorem}
A proof of Theorem~\ref{theorem1} is given in \ref{sec:app-pf-th}. 

\begin{theorem} \label{th:continuous-mapping}
The left side of \eqref{eq:generator-condition} implies that 
\begin{align} \label{eq:CTRW-limit}
\lim_{c \to \infty} X^{(c)}_t = X_t,
\end{align}
where $X^{(c)}_t$ is as in \eqref{eq:CTRW-trajectory}, and
\begin{align} \label{eq:subordination}
X_t = Y_{E(t)} \text{ where } E(t) = \inf\{u: Z_u > t\}.
\end{align}
The limit in \eqref{eq:CTRW-limit} is in the sense of stochastic processes with respect to the Skorokhod $J_1$ topology \cite{Whitt2010}. 
\end{theorem}

A proof of Theorem \ref{th:continuous-mapping} is given in \ref{sec:app-pf-th2}.
Putting together the two theorems and result \eqref{eq:generator-condition}, 
we see that conditions \eqref{eq:cond1}--\eqref{eq:cond4} are sufficient for the 
convergence \eqref{eq:CTRW-limit}.

\begin{example} \label{example}
Consider the kernel 
\begin{align}
\label{eq:example-kernel}
K^{(c)}(y,w|x,s) 
= \mathcal N\left(y | b(x,s+w)/c, a(x,s+w)/c\right)
\psi^{(c)}(w|x)
\end{align}
where $\mathcal N( \cdot | \mu, \sigma^2)$ denotes a Gaussian probability density with 
mean $\mu$ and variance $\sigma^2$, and where $\psi^{(c)}(w|x)$ is related to 
the spatially varying tempered stable L\'evy measure
\eqref{eq:varying-tempered-stable-levy-measure} as follows: 
Write 
\begin{align}
\overline \psi^{(c)}(w|x) = \int\limits_w^\infty \psi^{(c)}(w'|x)\,dw' 
\quad \text{ and } \quad 
\overline \nu(w|x) = \int\limits_w^\infty \nu(w'|x)\,dw'
\end{align}
for the tail functions of $\psi^{(c)}(w|x)$ and $\nu(w|x)$, and let 
\begin{align} \label{eq:def-psi}
\overline \psi^{(c)}(w|x) = 1 \wedge \left . \left( 
  c^{-1} 
  \overline \nu\left( w - \frac{d(x)}{c}\right| x
\right) \right), \quad w > 0
\end{align}
where we use the convention that $\overline \nu(w|x) = \infty$ if $w \le 0$. 
Then $\overline \psi^{(c)}(w|x)$ is decreasing and satisfies $\overline \psi^{(c)}(0|x) = 1$ and $\overline \psi^{(c)}(\infty|x) = 0$, so is hence the tail function of a probability density $\psi^{(c)}(w|x)$. 

The CTRW corresponding to \eqref{eq:example-kernel} has tempered heavy-tailed Pareto waiting times with spatially varying tail parameter $\beta(x)$ and tempering parameter $\theta(x)$. Its jumps are Gaussian with mean and variance equal to $b$ and $a$, evaluated at the current location and the time of jump. 
Using \eqref{eq:to-delta}, it is straightforward to check conditions \eqref{eq:cond1}--\eqref{eq:cond2}. 
Finally, using 
\begin{align}
\mathcal N\left(y | b(x,s+w)/c, a(x,s+w)/c\right) \to \delta_{(0,0)}(y), 
\quad c \to \infty,
\end{align}
\eqref{eq:cond3} follows from \eqref{eq:local-psi} and 
\eqref{eq:cond4} from \eqref{eq:non-local-psi}.
More details are in \ref{subsec:calc-for-example}.
Finally, we note that the construction of waiting time densities 
$\psi^{(c)}(w|x)$ via 
\eqref{eq:def-psi} from the tail function $\overline \nu(w|x)$ of a 
L\'evy measure satisfies \eqref{eq:cond3}--\eqref{eq:cond4} for \emph{any} 
L\'evy measure, not just 
\eqref{eq:varying-tempered-stable-levy-measure}.
\end{example}

\subsection{Remarks on spatio-temporal scaling parameters} \label{sec:scaling}

If the waiting time distribution $\psi^{(c)}(w|x)$ in 
\eqref{eq:example-kernel} has a finite mean $\tau$, a scaling 
$\tau \sim 1/c$ would yield a positive value for $d(x)$ in 
\eqref{eq:cond3} and a vanishing L\'evy measure in \eqref{eq:cond4}.
If the tails of $\psi^{(c)}(w|x)$ fall off as $w^{-1-\beta}$
(not depending on $x$), 
for instance
$\overline \psi^{(c)}(w|x) = (1+w/\tau)^{-\beta}$,
where 
$\beta \in (0,1)$, then the temporal scale $\tau$ needs to satisfy the 
scaling $\tau \sim c^{-1/\beta}$ to yield a meaningful limit in both 
\eqref{eq:cond3} and \eqref{eq:cond4} \cite[Section 3.7]{MeerschaertSikorskii}. 

If $\nu(w|x)$ has a variable
exponent $\beta(x)$ as in \eqref{eq:varying-tempered-stable-levy-measure},
it is no longer possible to satisfy the requirement \eqref{eq:cond4} via a 
global scaling relation $\tau \sim c^{-1/\beta}$.
For example, let the waiting times be as defined in \eqref{eq:STJK-example}.
The reader may check easily that for any choice of constant 
$\beta \in (0,1)$, 
the left side of \eqref{eq:cond4} will diverge or vanish, depending 
on whether $\beta(x)$ in $\nu(w|x)$ is smaller or larger than $\beta$.
Instead, the \emph{local} scaling relation
\begin{align}
  \tau(x) \sim c^{-1/\beta(x)}, \quad c \to \infty
\end{align}
needs to hold for \eqref{eq:cond3}--\eqref{eq:cond4} to be satisfied. 
In classical derivations of the FFPE (e.g.\ \cite{BMK00}), this is 
problematic, because the fractional diffusion coefficient can no longer be 
defined as the (global) limit of $\chi^2 / \tau^\beta$. 
The bivariate Langevin process $(Y_u^{(c)}, Z_u^{(c)})$, however, converges 
to a meaningful limit, and hence so does the CTRW $X_t^{(c)}$ (by Theorem 
\ref{th:continuous-mapping}).
Moreover, as discussed below, the distribution of the CTRW limit $X_t$ is 
the unique solution of the VOFFPE \eqref{eq:FFPE-with-scale}. 

For finite variance jumps as in the examples \eqref{eq:STJK-example}
and \eqref{eq:example-kernel}, 
the scaling $\chi^2 \sim 1/c$ is necessary to give a meaningful limit 
in equations \eqref{eq:cond1}--\eqref{eq:cond2}. 
For infinite variance jumps, where the tails of the jump probability 
density fall off as $x^{-\alpha-1}$ with $\alpha \in (0,2)$, 
the scaling $\chi^\alpha \sim 1/c$ is meaningful, see e.g.\ 
\cite[Section 3.8]{MeerschaertSikorskii}. 
This article only considers finite variance jumps; 
in order to apply to infinite variance jumps, the right side of \eqref{eq:cond4} 
needs to be replaced by 
$\iint g(z,w) K(z,w|x)\,dz\,dw$,
where for each $x$, $(z,w) \mapsto K(z,w|x)$ is a L\'evy density in 
$\spctim$ (see e.g. Equation (3.2) in \cite{BMMST}).

\section{Fokker-Planck Equations with Memory}
\label{sec:FFPE}

In \cite{BaeumerStraka16}, FFPEs for CTRW limit processes $X_t$ were derived
from the underlying infinitesimal generator $\mathcal A$ of the bivariate
process $(Y_u, Z_u)$.
Unlike the infinitesimal generator, the FFPE is expressed in the adjoint
space with ``forward'' variables $(y,t)$ (interpreted as the endpoint and
ending time of a trajectory) as opposed to the ``backward'' variables $(x,s)$
(i.e.\ the starting point and starting time; see e.g.\ 
\cite[Section 3.5.3]{Applebaum}).
The adjoint of the infinitesimal generator $\mathcal A$ in \eqref{eq:inf-gen}
is
\begin{align}
\label{eq:FPop}
\mathcal A^* g(y,t)
&= -\frac{\partial }{\partial y}\left[b(y,t) g(y,t)\right]
+\frac{1}{2}\frac{\partial^2 }{\partial y^2}\left[a(y,t) g(y,t)\right]
\\
\label{eq:Dop}
&-\frac{\partial }{\partial t}[d(y) g(y,t)]
+ \int\limits_{w > 0} \left[g(y, t-w)-g(y, t)
\right] \nu(w | y) \, dw
\\ \notag
&= [\mathcal L^* + \mathcal D^*] g(y,t)
\end{align}
where $\mathcal L^*$ (the Fokker-Planck operator) is the spatial differential
operator on the right-hand side of \eqref{eq:FPop} and $\mathcal D^*$ is the
temporal integro-differential operator in \eqref{eq:Dop}.
Since the infinitesimal generator decomposes as
$\mathcal A = \mathcal L + \mathcal D$, the probability distribution $P(y,t)$
of $X_t$ satisfies the Fokker-Planck equation in \cite[(5.2)]{BaeumerStraka16}:
\begin{align} \label{eq:FFPE}
\frac{\del P(y,t)}{\del t} = \mathcal L^* \left[ \frac{\partial}{\partial t}
\int_0^t P(y,t-s) V(y,s)\,ds \right] + h(y,t).
\end{align}
Here $h(y,t)$ is an initial condition (typically $h(y,t) = \delta(y) \delta(t)$)
and $V(y,t)$ is a spatially dependent memory kernel defined via its Laplace
transform\footnote{To see that this definition is equivalent to Proposition 5.1
in \cite{BaeumerStraka16}, note that $H(y,w) \equiv \overline \nu(w|y) = \int_w^\infty \nu(w|y)$,
and apply integration by parts.}
\begin{align} \label{eq:LT-renewal-measure}
\begin{split}
\hat V(y,\lambda) = \int_0^\infty e^{-\lambda t} V(y,t)\,dt
&= \frac{1}{d(y)\lambda + \int_0^\infty (1-e^{-\lambda w})
\nu(w | y)\,dw}
= \frac{1}{\lambda} \, \frac{1}{d(y) + \hat{\overline{\nu}}(\lambda|y)}.
\end{split}
\end{align}

\paragraph{Renewal measure}
As can be seen from the infinitesimal generator \eqref{eq:inf-gen},
if $x \in \spc$ is held fixed, the temporal component $Z_u$ of the Langevin
process
is a subordinator with infinitesimal generator
\begin{align}
\mathcal D f(x,s) = d(x) \frac{\partial }{\partial s} f(x,s)
+ \int_0^\infty [f(x, t+w) - f(x,s)] \nu(w|x)\,dw.
\end{align}
By the L\'evy-Khintchine theorem (see e.g.\ \cite{Bertoin04} for its form
for subordinators), its Laplace transform is
$\langle e^{-\lambda Z_u(x)} \rangle = e^{-u \Upsilon(\lambda|x)}$, where the
so-called L\'evy symbol $\Upsilon(\lambda|x)$ is
\begin{align} \label{eq:symbol}
\Upsilon(\lambda|x) = d(x) \lambda + \int_0^\infty (1-e^{-\lambda w})
\nu(w|x)\, dw.  
\end{align}
Moreover, $1/\Upsilon(\lambda| x) = \hat V(x, \lambda)$ is the Laplace
transform of the renewal measure density $V(x,s)$ of $Z_u(x)$, which measures
the expected occupation time of $Z_u(x)$ (see e.g.\ 
\cite[Eq.(15)]{MMMPS12}, or \cite{Bertoin04}):
\begin{align}
\int_a^b V(x,s)\, ds = \left \langle \int_0^\infty \1(a < Z_u(x) < b)\,du
\right \rangle,  
\end{align}
where $\langle\,\rangle$ denotes expectation or the ensemble average over all
$(Y_u,Z_u)$ trajectories.
Note that the interval $(a,b)$ is a \emph{physical} time interval, and the 
(random) amount of time spent in it by $Z_u$ is \emph{auxiliary} time, akin to 
the number of steps that the walker takes in the time window $(a,b)$. 
Hence $V(y,t)$ measures unit auxiliary time per unit physical time. The brackets
in \eqref{eq:FFPE} are hence a quantitiy of unit
``probability $\times$ auxiliary time / physical time'',
and the Fokker-Planck operator of unit ``1 / auxiliary time'',
thus giving \eqref{eq:FFPE} a dimensionally consistent interpretation.

\begin{remark}
The temporal drift coefficient $d(x)$ in \eqref{eq:symbol} has a more
straightforward representation in a less general case:  
Assume that 
$\nu(w|y) \equiv \nu(w)$ and $d(x) \equiv d$ do not depend on $x$, 
and that drift $b(x,t)$ and diffusivity $a(x,t)$ are constant in $t$.  
Then by comparison of Laplace transforms, it follows that \eqref{eq:FFPE}
is equivalent to 
\begin{align} \label{eq:distributed-order}
d\,\frac{\del }{\del t} P(y,t) 
+ \int_0^t \left[\frac{\del }{\del t} P(y,t-s)\right] 
\overline{\nu}(s)\,ds = \mathcal L^* P(y,t) 
+ H(y,t)
\end{align}
where $H(y,t) = dh(y,t) + \int h(y,t-s) \overline \nu(s)\,ds$.
Three special cases are of interest:
\begin{itemize}
\item
If $d=1$ and $\nu(w) \equiv 0$, then \eqref{eq:distributed-order} 
reduces to the (non-fractional) Fokker-Planck equation. 
\item
If $d=0$ and $\nu(w)$ is as in \eqref{eq:stable-levy-measure}, then 
\eqref{eq:distributed-order} becomes the governing equation for 
Hamiltonian chaos in the sense of \cite{Zaslavsky94}, see also 
Eq.(24) in \cite{Baeumer2001}. 
\item
If $d \ge 0$ and $\nu(w)$ is as in 
\eqref{eq:stable-levy-measure}, then 
\eqref{eq:distributed-order} is a special case of a ``distributed order'' 
FFPE \cite{Sandev2015}, where the distribution over the orders 
is a mixture of the two orders $1$ and $\beta$.
\end{itemize}
\end{remark}

\paragraph{Inhomogeneous scaling}
For spatially varying tempered subdiffusion, i.e.\ when $\nu(w|x)$ has the form 
\eqref{eq:varying-tempered-stable-levy-measure}, 
the Laplace transform \eqref{eq:LT-renewal-measure} of $V(y,t)$ takes the form
\begin{align}
\hat V(y,\lambda) =
\frac{1}{d(y) \lambda + (\lambda + \theta(y))^{\beta(y)} - \theta(y)^{\beta(y)}}.
\end{align}
We are only able to invert the Laplace transform in the case $d \equiv 0$ and
$\theta \equiv 0$, whence $V(y,t) = t^{\beta(y)-1} / \Gamma(\beta(y))$; 
for the remainder of this article, we restrict ourselves to this case.
To make the spatio-temporal scaling in the dynamics explicit, we introduce the time units
$T_0$ for physical time, $A_0$ for auxiliary time and $L_0$ for length, 
and recall that $V(y,t)$ measures unit auxiliary time per unit physical 
time:
$$V(y,t) = \frac{(t/T_0)^{\beta(y)-1}}{\Gamma(\beta(y))}\, \frac{A_0}{T_0}
= \frac{A_0}{T_0^{\beta(y)}} \, \frac{t^{\beta(y) - 1}}{\Gamma(\beta(y))}$$
and hence convolution with $V(y,t)$ becomes the Riemann-Liouville
fractional derivative of spatially varying order $1-\beta(y)$, 
\begin{align}
  \frac{\partial}{\partial t} \int_0^t P(y,t-s) V(y,s)\,ds
  = {_0D}_t^{1-\beta(y)} P(y,t) \, \frac{A_0}{T_0^{\beta(y)}}
\end{align}
(of unit $A_0 / T_0$).
The Fokker-Planck operator is
\begin{align}
\mathcal L^* g(y,t)
&= -\frac{\partial }{\partial y}[b(y,t) g(y,t)]
+\frac{\partial^2 }{\partial y^2}[a(y,t) g(y,t)],
\end{align}
and the FFPE \eqref{eq:FFPE} then reads
\begin{align} \label{eq:FFPE-with-scale}
\frac{\partial P(y,t)}{\partial t}
= \frac{\partial^2}{\partial y^2} \left[a_{\beta(y)}(y,t)
  \, {_0 D}_t^{1-\beta(y)} P(y,t)\right]
  - \frac{\partial}{\partial y} \left[b_{\beta(y)}(y,t)
  \, {_0 D}_t^{1-\beta(y)} P(y,t)\right]
  + h(y,t),
\end{align}
where 
\begin{align} \label{eq:generalized-coefficients}
  a_{\beta(y)}(y,t) &= a(y,t) \frac{A_0}{T_0^{\beta(y)}}, 
  & 
  b_{\beta(y)}(y,t) &= b(y,t) \frac{A_0}{T_0^{\beta(y)}}
\end{align}
are the generalized diffusivity and drift, with units 
$L_0^2/T_0^{\beta(y)}$ and $L_0/T_0^{\beta(y)}$, respectively. 

If the fractional exponent and the diffusivity are constant, i.e.\ 
$\beta(y) \equiv \beta \in (0,1)$ and $a(y,t) \equiv 1$ (with unit $L_0^2/A_0$),
we retrieve the well-known FFPE (see e.g.\ \cite[Eq.(3)]{HLS10PRL})
with fractional diffusion constant $a_\beta = L_0^2 / T_0^\beta$.
(Note that in \cite{HLS10PRL}, the space- and time-dependent fractional 
drift $b(y,t) / T_0^\beta$ coefficient is expressed as a space- and 
time-dependent force $F(y,t)$, divided by a fractional friction parameter 
$\eta_\beta$.)
We remind the reader that in our Langevin framework, $a(y,t)$ and 
$b(y,t)$ need not be constant, but can be any bounded differentiable 
functions with bounded derivative. 

The auxiliary time scale parameter 
$A_0$ cancels in \eqref{eq:generalized-coefficients}, since 
$a(y,t)$ is measured in $L_0^2 / A_0$ and $b(y,t)$ is measured in $L_0 / A_0$,
for some unit of length $L_0$. 
Thus the choice of $A_0$ \emph{has no effect on \eqref{eq:FFPE-with-scale}}.
The interpretation is that the
Fokker-Planck dynamics are independent of the ``speed'' of the Langevin process
$(Y_u, Z_u)$.  This is to be expected; to see why, first observe that
multiplying $\mathcal A^*$ by a constant $C > 0$ is equivalent to dividing 
$A_0$ by $C$ and uniquely corresponds to the
Langevin process $(Y_{Cu}, Z_{Cu})$. 
Now note that the graphs of the two processes, i.e.\ the sets
\begin{align*}
\{(u, Y_u, Z_u): u \ge 0\} \quad \text{ and } \quad 
\{(u, Y_{Cu}, Z_{Cu}): u \ge 0\}
\end{align*}
have different distributions, whereas the \emph{images}
\begin{align*}
\{(Y_u, Z_u): u \ge 0\} \quad \text{ and } \quad 
\{(Y_{Cu}, Z_{Cu}): u \ge 0\},
\end{align*}
i.e.\ the points traversed by $(Y_u,Z_u)$ resp.\ $(Y_{Cu}, Z_{Cu})$, 
have identical distributions. 
Since the CTRW scaling limit is uniquely constructed from the image via 
\eqref{eq:subordination},
as opposed to from the graph,
it must have the same distribution for both Langevin processes.

The choice of time scale is intrinsic in the generalized diffusivity and drift. 
If the unit of time $T_0$ is changed, e.g.\ increased by a factor of $10$, 
then according to \eqref{eq:generalized-coefficients}, 
$a_{\beta(y)}(y,t)$ and $b_{\beta(y)}(y,t)$ must be divided by 
$10^{\beta(y)}$. 
Then the CTRW limit $X_{10t}$ at time $10 t$ with the new coefficients must 
have the same distribution as the CTRW limit $X_t$ at time $t$ with the old 
coefficients. 
We check if this is the case in the next section.

\section{Monte Carlo approximation of an FFPE solution}
\label{sec:examples}

In \cite{Korabel2010}, the authors study an interface problem, where for 
$x<0$ the medium is more subdiffusive ($\beta = 0.3$) than for $x>0$ 
(where $\beta = 0.75$).  Even if a positive drift is applied at the 
interface, eventually all particles end up in the left half, even though 
the first moment $\langle x \rangle$ of the solution is always increasing. 
The reason for this ``paradoxical'' behaviour is as follows: the random 
walk on the infinite line is recurrent, i.e.\ each point is due to be 
visited an infinite number of times. Since the heavier-tailed waiting 
times in the left dominate the waiting times in the right, as $t \to 
\infty$ particles tend to be trapped in the left. Compared to the left, 
particles on the right are moving freely, and hence spread out further 
towards $+\infty$, hence the mean $\langle x \rangle$ is ever increasing. 

\begin{figure}
  \centering
  \includegraphics{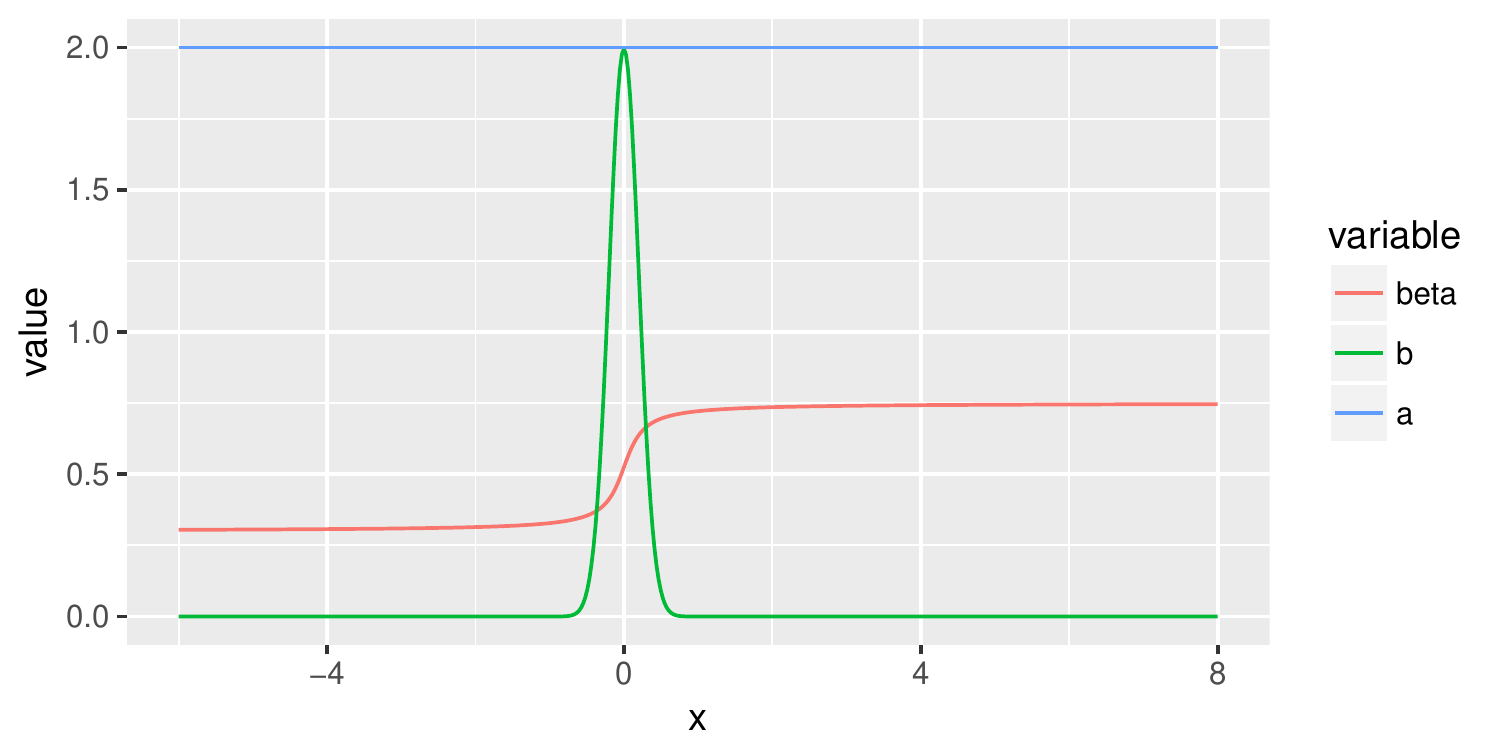}
  \caption{\label{fig:beta_b}Coefficients imitating the interface problem from \cite{Korabel2010}.}
\end{figure}

We aim to reproduce this interface problem, and 
let the anomalous exponent be $\beta(y) = 0.525 + 0.45 * \arctan(y/0.2)/ \pi$, 
which interpolates smoothly between $\lim_{y \to -\infty} \beta(y) = 0.3$ 
and $\lim_{y \to +\infty} \beta(y) = 0.75$, yet creates a sharp change at $y=0$, see Figure \ref{fig:beta_b}.
For the drift $b(y,t)$, we use the density of a Gaussian distribution with 
mean $0$ and standard deviation $0.2$, to approximate smoothly a 
delta function pulse at $0$.
The diffusivity $a(y,t) = 2$ is constant. 

We check whether equation \eqref{eq:FFPE-with-scale} is consistent if 
measured in two 
different time units, $T_0 = 1$ (second) and $T_0 = 10$ (seconds).
If it is, then the probability densities of $X_t$ at the times 
$1/8, 1, 8, 64$ for $T_0 = 1$ must be the same as the probability 
densities at the times $10/8, 10, 80, 640$ for $T_0 = 10$.
Observe that changing $T_0 = 1$ to 
$T_0 = 10$ is equivalent to
dividing both $a(x,t)$ and $b(x,t)$ by $10^{\beta(x)}$. 

\begin{figure}[h]
\centering
\includegraphics{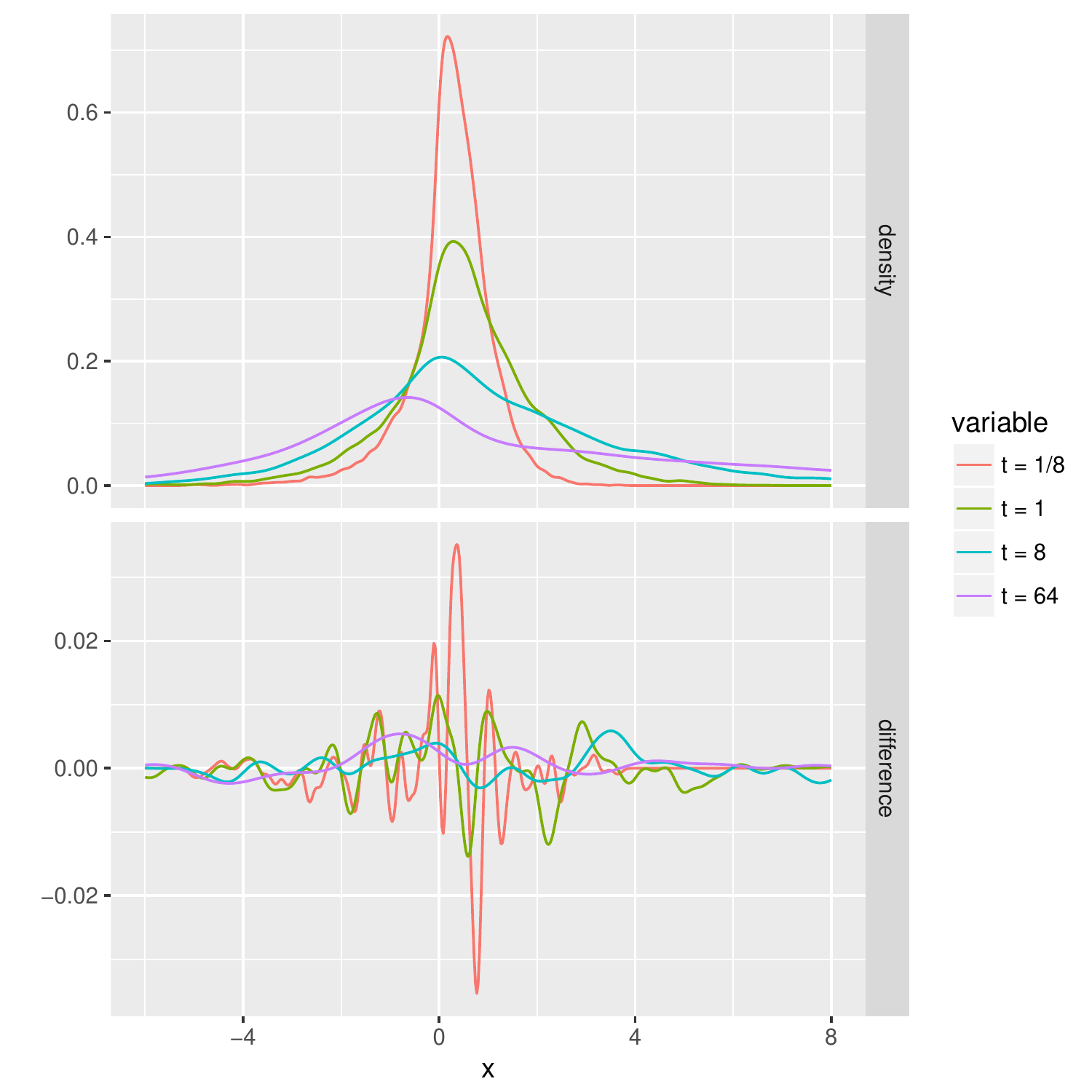}
\caption{\label{fig:comparison}Comparison of the dynamics of variable fractional order, at two different time scales $T_0 = 1$ and $T_0 = 10$. The upper panel shows the probability densities for $T_0 = 1$ at different times. The lower panel shows the difference between probability densities for $T_0 = 10$ and for $T_0 = 1$.}
\end{figure}

We have simulated 10,000 trajectories with the two coefficients scenarios above, 
at the scale 
parameter $c = 200$ (view the R-code at \cite{var-order-MC}).
The value $c = 200$ was chosen so that the mean number of steps at the smallest time 
$t = 1/8$ was above $30$. The trajectories were evaluated at the times mentioned
above, and a kernel density estimator (with automated bandwidth selection) was applied
to the empirical distributions at each time, thus producing a Monte Carlo 
estimate of the probability distributions.
The probability density estimates for $T_0$ are shown in the upper panel 
in Figure \ref{fig:comparison}. 
The lower panel shows the difference between the densities in the two 
coefficient scenarios. 
This difference is seen to be falling within the expected range of Monte Carlo 
fluctuations, and is uniformly below $0.04$ at all times. 
This fact supports the consistency of the VOFFPE over varying time scales 
\eqref{eq:FFPE-with-scale}. 

\section{Conclusion}

In this article, we have employed the theory of stochastic differential equations (SDEs) with
L\'evy noise for the general representation of CTRW limit processes with spatially varying memory kernels.  The theory allows for memory kernels of any variety, 
the only restriction on the type of memory kernel being the L\'evy measure condition \eqref{eq:Levy-condition}.  The memory kernel can then be meaningfully varied in space, as long as certain technical assumptions on the SDE coefficient functions (local linear growth \& local Lipschitz continuity) are maintained.  

Our main motivation was to give a rigorous treatment of CTRW scaling limits and FFPEs with spatially varying fractional order, and hence we have focussed on situations of this type. The discussion surrounding variable tempering and temporal drift parameters we have left for future work.  
Here, we have given a derivation which identifies the unique solutions of variable-order FFPE as probability distributions of CTRW scaling limits. 
Moreover, we have constructed a sequence of CTRW processes which converges to the solution of a variable-order FFPE.  We have used this approach to approximate a solution to a variable-order FFPE via Monte Carlo simulation.

The fractional exponent $\beta$ has traditionally been used to define a global fractional diffusion coefficient, of dimension Length$^2$ / Time$^\beta$.  For spatially varying $\beta(x)$, such a global diffusion coefficient no longer exists.  
Instead, a factor $T_0^{-\beta(x)}$ arises in the drift and diffusion 
coefficients of the variable order FFPE, where $T_0$ corresponds to the applicable unit of time.  We have given a dimensional analysis which shows that the variable order FFPE is dimensionally consistent.  Finally, our simulations confirm that although the drift and diffusivity parameters seemingly depend on the chosen time scale $T_0$ and thus yield different governing FFPEs, the solutions of these different FFPEs are identical and hence that the theory is consistent. 

Currently, there exist few numerical methods for the solution of FFPEs with memory kernels which vary in space, although such FFPEs may well be useful to the physics community, as discussed in the introduction.  The algorithm from \cite{Gill2016} may be extended to the spatially inhomogeneous setting, an idea which we will follow up on in future work.

\subsection*{Acknowledgements}
Peter Straka was supported by the Australian Research Council via a 
Discovery Early Career Researcher Award (DECRA) DE160101147. 
The author thanks Christopher Angstmann, Gurtek Gill, 
Sergei Fedotov, Bruce Henry, James Nichols and Nickolay Korabel for stimulating discussions which 
significantly improved the presentation of this article. 

\section*{References}

\bibliography{varyExp}

\appendix

\section{Appendix}

\subsection{Lipschitz and growth conditions on the coefficients} 
\label{subsec:Lip-gro}
After truncation\footnote{Truncation of jumps has no effect on the existence and 
uniqueness of solutions to Stochastic Differential Equations \cite{Applebaum}.} 
at a number $l > 0$, the variable tempered stable L\'evy 
measure 
\eqref{eq:varying-tempered-stable-levy-measure}
has the same form as the function $g$ in Equation (3.1) of 
\cite{Tsuchiya1992}.  
We note that we may ignore $t$ in their notation (our $u$), since we only 
consider Langevin processes with dynamics that are homogeneous in $u$.
We may also drop $\theta$ in their notation, as our L\'evy noise occurs only 
in the one direction $(0,1)$. 
Then identifying $n(x;\rho) \equiv \exp(\theta(x) w)$, 
$\alpha(x) \equiv \beta(x)$, $\beta(x) \equiv 0$, 
we have $g(x;d\rho) \equiv \nu(w|x)\,dw$. 
The condition $\mathcal L_x([0,\infty), R^d, S, (0,l])$ on $n$, $\alpha$ and 
$\beta$ is satisfied if we assume that $\theta(x)$ and $\beta(x)$ are 
continuously differentiable with bounded derivative. 
If $\theta(x)$ and $\beta(x)$ are also bounded, then all four sub-conditions of 
Condition (II) are easily seen to be satisfied.  
Hence (s. top of page 111) part (ii) of Condition (I) is satisfied. 
Part (i) of Condition (I) is trivially fulfilled since 
$y(x; \rho) = \rho \equiv w$ is the identity function. 
Hence Theorem 2.1 applies, which means that $F(w|x)$ satisfies the Lipschitz 
and growth conditions from Theorem 1.1. 
Since we assume that the remaining coefficients are bounded and differentiable 
with bounded derivative, they also satisfy the Lipschitz and growth conditions, 
and hence the Langevin equations \eqref{eq:SDEY}--\eqref{eq:SDEZ} have a unique 
solution $(Y_u, Z_u)$.

\subsection{Proof of Theorem \ref{theorem1}}
\label{sec:app-pf-th}

Assume that conditions \eqref{eq:cond1}--\eqref{eq:cond4} all hold. 
Split the domain of integration in \eqref{eq:jump-process-generator} into a local part $L = \{(y,w): |y| \le \varepsilon, \, 0 < w \le \varepsilon\}$ and its complement $L^\complement$. On $L$, we do a Taylor expansion and calculate
\begin{align*}
&c \iint\limits_L [f(x+y, s+w) - f(x,w)] K^{(c)}(y,w|x,s)\,dy\,dw
\\
= &\iint\limits_L \left[y f_x(x,s)
+ w f_s(x,s)
+ \frac{1}{2} y^2 f_{xx}(x,s) 
+ yw f_{xs}(x,s) 
+ w^2 f_{ss}(x,s)
+ o(|y|^2 + |y||w| + |w|^2)\right]
\\
& c K^{(c)}(y,w | x,s) \,dy\,dw
\\
= &f_x(x,s) \iint\limits_L y \,cK^{(c)}(y,w|x,s)\,dy\,dw
+ f_s(x,s) \iint\limits_L w \,cK^{(c)}(y,w|x,s)\,dy\,dw
\\
+ &f_{xx}(x,s) \iint\limits_L y^2 \,cK^{(c)}(y,w|x,s)\,dy\,dw
+ f_{xs}(x,s) \iint\limits_L yw \,cK^{(c)}(y,w|x,s)\,dy\,dw
\\
+ &f_{ss}(x,s) \iint\limits_L w^2 \,cK^{(c)}(y,w|x,s)\,dy\,dw
+ o(1)
\end{align*}
where $o(1)$ is a term that tends to $0$ as $\varepsilon \downarrow 0$. 
For the $yw$-term, we find 
\begin{align*}
\left|\iint\limits_L yw \,cK^{(c)}(y,w|x,s)\,dy\,dw \right|
\le \iint\limits_L |y|w \,cK^{(c)}(y,w|x,s)\,dy\,dw
\\
\le \varepsilon \iint\limits_L w \,cK^{(c)}(y,w|x,s)\,dy\,dw
\to \varepsilon [d(x) + o(1)],
\end{align*}
and the same bound holds for the $w^2$-term. 
We add the above and the integral over the complement $L^\complement$ and let $c \to \infty$, to get
\begin{align*}
f_x(x,s) b(x,s) + o(1) + f_s(x,s) d(x) + o(1) + f_{xx}(x,s) a(x,s) + o(1) 
+ 2\varepsilon [d(x) + o(1)]
\\
+ \iint\limits_{L^\complement} [f(x, s+w) - f(x,s)] \nu(w|x)\,dw.
\end{align*}
As $\varepsilon$ was arbitrary to begin with, we let $\varepsilon \downarrow 0$, 
and the above equals $\mathcal A f(x,s)$. 

\subsection{Proof of Theorem \ref{th:continuous-mapping}}
\label{sec:app-pf-th2}

Write $(y,z)$ for a generic path in $\spctim$, which is such that 
$z$ is unbounded and non-decreasing.
By Proposition 2.3 in \cite{StrakaHenry}, the following path mapping is 
continuous on the subset of all paths $(y,z)$ which are such that $z$ is 
strictly increasing: 
$(y,z) \mapsto (y_- \circ z^{-1}_-)_+$, which composes the left-continuous 
version of $y$ with the left-continuous version of the inverse 
\begin{align}
  z^{-1}(t) = \inf\{u: z(u) > t\}
\end{align}
and changes the resulting composition back to right-continuous again. 
This path mapping maps the continuous time Markov chain paths 
$u \mapsto (Y^{(c)}_u, Z^{(c)}_u)$ to the CTRW paths 
$t \mapsto X^{(c)}_t$ (Lemma 3.5 in \cite{StrakaHenry}). 
Since the limiting process $(Y_u, Z_u)$ has unbounded and strictly increasing 
$Z_u$, with probability $1$ its samples paths lie in the set of continuity of 
the above path mapping. 
The continuous mapping theorem \cite{Billingsley1968} applies, saying that 
$\lim_{c \to \infty}(Y^{(c)}_u, Z^{(c)}_u) = (Y_u, Z_u)$ with respect to $J_1$
implies
$\lim_{c \to \infty} X^{(c)}_t = (Y_- \circ Z^{-1}_-)_{t+}$ with respect to $J_1$. 
Note that $Y_u$ has continuous paths, and so does $Z^{-1}_u$ because 
$Z_u$ is strictly increasing. But then 
$(Y_- \circ Z^{-1}_-)_{t+} = Y \circ Z^{-1} (t) = Y_{E(t)}$.

\subsection{Calculation for Example \ref{example}}
\label{subsec:calc-for-example}

First, we note that 
\begin{align}
  \overline \nu(w|x) \sim w^{-\beta(x)} / \Gamma(1-\beta(x)), 
  \quad w \downarrow 0, 
\end{align}
and thus the inverse function ${\overline \nu}^{-1}(c|x)$ of 
$z \mapsto \overline \nu(z|x)$ satisfies
\begin{align}
  \overline \nu(z | x) \le c
  \Leftrightarrow z \ge {\overline \nu}^{-1}(c|x)
  \sim (\Gamma(1-\beta(x))c)^{-1/\beta(x)},
\end{align}
Hence from the definition \eqref{eq:def-psi} of $\overline \psi^{(c)}(w|x)$, 
\begin{align}
\label{eq:psi-c}
  c \psi^{(c)}(w|x) = \begin{cases}
  \nu(w | x) & \text{ if } w > C(x,c) 
  := {\overline \nu}^{-1}(c|x) + d(x)/c, 
  \\
  0 & \text{ if } 0 < w \le C(x,c). 
  \end{cases}
\end{align}
Since $C(x,c) \downarrow 0$ as $c \to \infty$, for any $w > 0$
\begin{align} \label{eq:non-local-psi}
c \psi^{(c)}(w|x) \to \nu(w|x),
\quad 
  c \overline \psi^{(c)}(w|x) \to \overline \nu(w|x), \quad c \to \infty. 
\end{align}

We also note that $\psi^{(c)}(w|x)$ is a probability density on the positive 
numbers.  Due to \eqref{eq:psi-c}, for any $w > 0$, and large enough $c$,
\begin{align}
  \psi^{(c)}(w|x) = \nu(w|x) / c \to 0, \quad c \to \infty. 
\end{align}  
Hence as $c \to \infty$, the probability mass of the density 
$\psi^{(c)}(w|x)$ must get concentrated near $0$, i.e.\ 
\begin{align} \label{eq:to-delta}
  \int_0^\infty f(w) \psi^{(c)}(w|x) \to f(0)
\end{align}
for any bounded continuous $f$.

Finally, for any $\varepsilon > 0$,
\begin{align}
\label{eq:local-psi}
\begin{split}
  c \int_0^\varepsilon w \psi^{(c)}(w|x) \,dw
  &= c \int_{C(x,c)}^\varepsilon w \psi^{(c)}(w|x) \,dw
  \\
  &= -c\left[w\overline \psi^{(c)}(w|x)\right]^\varepsilon_{C(x,c)}
  + c \int_{C(x,c)}^\varepsilon \overline \psi^{(c)}(w|x)\,dw
  \\
  &= -c\varepsilon \overline \psi^{(c)}(\varepsilon|x)
  + c C(x,c) \times 1
  + c \int_{C(x,c)}^\varepsilon \overline \psi^{(c)}(w|x)\,dw
  \\
  &\to \varepsilon \overline\nu(\varepsilon | x)
  + d(x) + \int_0^\varepsilon \overline \nu(w|x)\,dw
  \\
  &= d(x) + \mathcal O(\varepsilon^{1-\beta(x)}).
\end{split}
\end{align}

\end{document}